\documentclass[11pt,a4paper]{scrartcl}

\usepackage{CLICdp}
\usepackage{float}

\usepackage{CLICdp_definitions}

\RequirePackage{heppennames2}

\newcommand{\MeV}{\ensuremath{\text{MeV}}\xspace}
\newcommand{\GeV}{\ensuremath{\text{GeV}}\xspace}
\newcommand{\TeV}{\ensuremath{\text{TeV}}\xspace}
\newcommand{\tabt}[1]{\multicolumn{1}{c}{#1}}
\newcommand{\tabtt}[1]{\multicolumn{2}{c}{#1}}
\newcommand{\mH }{\ensuremath{m_{\PH}}\xspace}
\newcommand{\GH }{\ensuremath{\Gamma_{\PH}}\xspace}
\newcommand{\BR}{\ensuremath{BR}\xspace}

\newcommand{\gHZZ}{\ensuremath{g_{\PH\PZ\PZ}}\xspace}
\newcommand{\gHWW}{\ensuremath{g_{\PH\PW\PW}}\xspace}
\newcommand{\gHtt}{\ensuremath{g_{\PH\PQt\PQt}}\xspace}
\newcommand{\gHbb}{\ensuremath{g_{\PH\PQb\PQb}}\xspace}
\newcommand{\gHcc}{\ensuremath{g_{\PH\PQc\PQc}}\xspace}
\newcommand{\gHTauTau}{\ensuremath{g_{\PH\PGt\PGt}}\xspace}
\newcommand{\gHMuMu}{\ensuremath{g_{\PH\PGm\PGm}}\xspace}

\title{Updated CLIC luminosity staging baseline and Higgs coupling prospects}

\clicdpnote{2018}{002}

\date{26 November 2018}

\addauthor{A.~Robson\thanks{Also at CERN, Geneva, Switzerland}}{\institute{1}}
\addauthor{P.~Roloff}{\institute{2}}

\addinstitute{1}{University of Glasgow, United Kingdom}
\addinstitute{2}{CERN, Geneva, Switzerland}

\abstract{An updated luminosity staging baseline for CLIC is presented. 
  Assuming accelerator ramp-up and up-time scenarios that are
  harmonized with those of other potential future colliders, 
  CLIC will deliver $1\,\abinv$ at $\roots=380\,\GeV$, $2.5\,\abinv$ at $\roots=1.5\,\TeV$,
  and $5\,\abinv$ at $\roots=3\,\TeV$.  The complete programme will take 25--30 years.
  The baseline scenario for luminosity sharing between the two electron
  beam polarisation states is also discussed. 
  Updated Higgs coupling sensitivities are given for this new luminosity
  staging baseline.}

\graphicspath{ {./logos/}{./figures/} }

\begin{document}

\titlepage

\section{Introduction}

CLIC offers high-energy \epem\ collisions up to centre-of-mass energies of 3\,TeV \cite{clic-study}.
A rich programme of Higgs and top-quark physics is uniquely provided by the initial energy
stage around $\roots=380\,\GeV$; this is supplemented at the higher-energy stages
by increased precision in Higgs and top-quark physics, and further reach to beyond-Standard Model (BSM) effects. 

The revised luminosity baseline comes in the context of an initiative 
to harmonize the assumptions used by several future collider projects \cite{Bordry:2018gri},
and presents an opportunity to reassess the optimal staging scenario for physics reach
in light of recent studies.  
The effect of harmonizing the collider up-time and 
ramp-up assumptions, as well as modest adjustments to the time spent at each
CLIC energy stage, is to increase substantially the integrated luminosity that is
foreseen for CLIC.

The updated running assumptions and resulting integrated luminosities are described
in \autoref{sec:staging} of this note.  A baseline polarisation scenario is given in
\autoref{sec:polarisation}, and the resulting Higgs coupling sensitivities are
discussed in \autoref{sec:higgs}.

\section{Staging}
\label{sec:staging}

In the revised scenario, the assumed run period per year is 185 days.
This allows for a long shutdown
of around 120 days per year; start-up of around 30 days; and longer 
machine development and technical stops of around 30 days.  The assumption of 75\%
efficiency during the 185-day run period, to allow for faults and preventative
maintenance, results in a running time of $1.2 \times 10^7$\,s per year.
This is based on previous CERN experience and has been agreed with the FCC projects
as a common baseline.  It is less than the ILC scenario, which assumes running
for $1.6 \times 10^7$\,s per year (computed as 75\% of 8 months, i.e. running for 185 days) \cite{ILCrunningScenarios}.
However, it is more than what was foreseen in the previous CLIC baseline, which assumed
a running time of $1.08 \times 10^7$\,s per year \cite{staging_baseline_yellow_report}.

At the initial energy stage, $\roots=380\,\GeV$, the accelerator is expected to
ramp up to its nominal performance over the first three years of running. 
In this revised baseline, the delivered luminosity is assumed to be 10\%, 30\%,
and 60\% of the nominal luminosity in each of these ramp-up years, respectively. 
This ramp-up scenario matches that assumed by the ILC. 
The three-year ramp-up period will be followed by five years running at the nominal
luminosity, to deliver a total integrated luminosity of $1\,\abinv$.  This 
includes $100\,\fbinv$ taken in an energy scan around the \ttbar production threshold.

For the subsequent energy stages, 1.5\,TeV and 3\,TeV, the accelerator is
expected to reach its nominal performance after two years, and a ramp-up 
of 25\% and 75\% of the nominal luminosity in the first two years of
each stage is assumed.  This also matches the ILC scenario.

At $\sqrt{s}=1.5\,\TeV$, the ramp-up will be followed by five years of running at the
nominal luminosity, yielding a total integrated luminosity of 2.5\,\abinv.
At $\sqrt{s}=3\,\TeV$, the ramp-up will be followed by six years of running at the
nominal luminosity, for a total integrated luminosity of 5\,\abinv.

At each stage the collider is run for slightly longer 
in the revised scenario, compared with the previous
scenario \cite{staging_baseline_yellow_report}.
Running for longer at each stage, when the collider
has reached its nominal luminosity, both 
takes advantage of the machine's full design performance, and
is more consistent with the cost and time requirements of staging the machine's construction.

\begin{figure}
\begin{center}
\includegraphics[width=10cm]{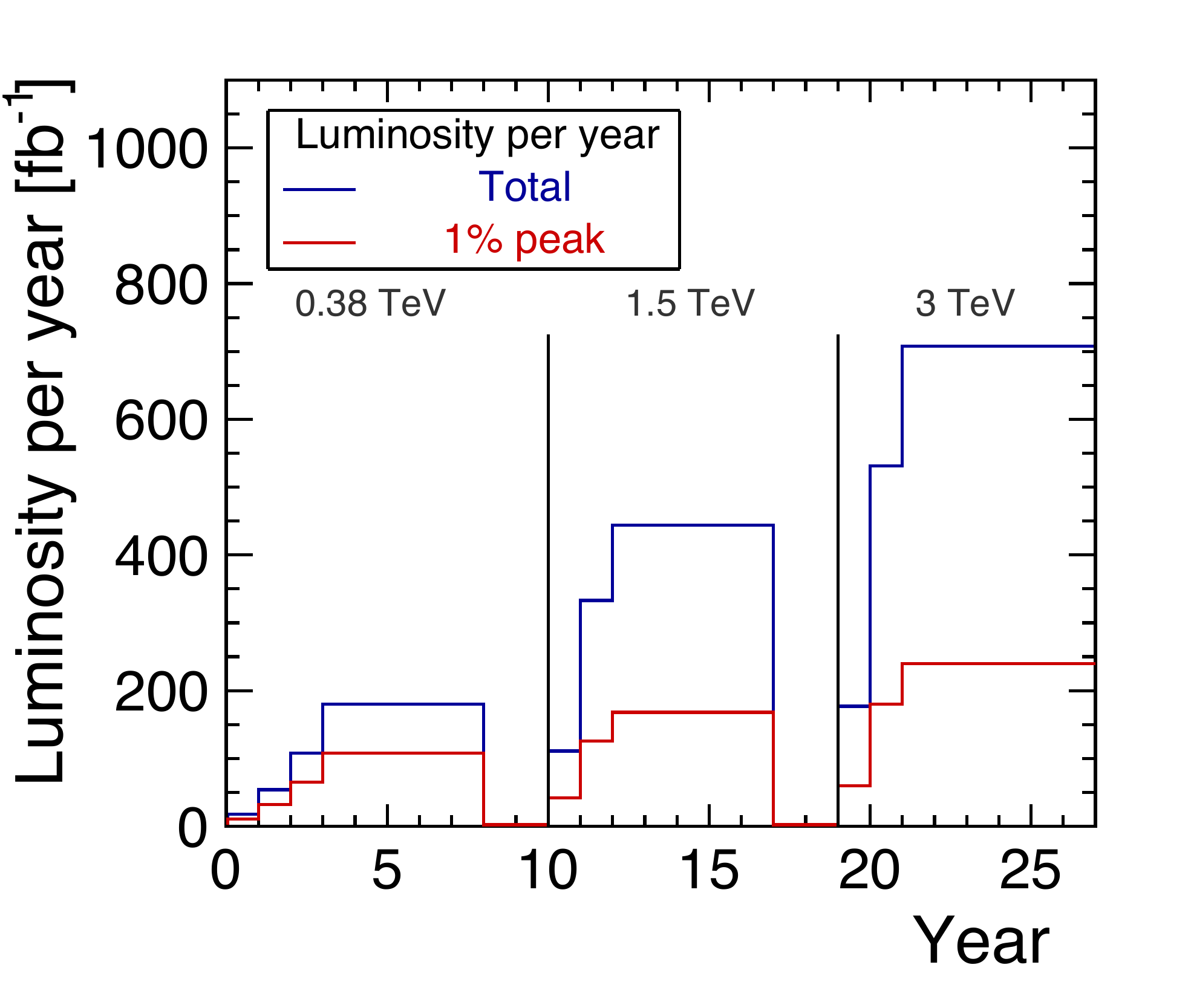}
\caption{Luminosity per year in the updated CLIC staging scenario.  For the first stage the luminosity
  ramp-up is three years (10\%, 30\%, 60\%) and for the subsequent stages two years (25\%, 75\%).
  As a result of beamstrahlung the CLIC beam spectrum has a low-energy tail, so both the total
  luminosity per year, and the luminosity collected above 99\% of the nominal \roots\ (labelled `1\% peak'),
  are shown.  \label{fig:clicstaging1}}
\end{center}
\end{figure}

\begin{figure}
\begin{center}
\includegraphics[width=10cm]{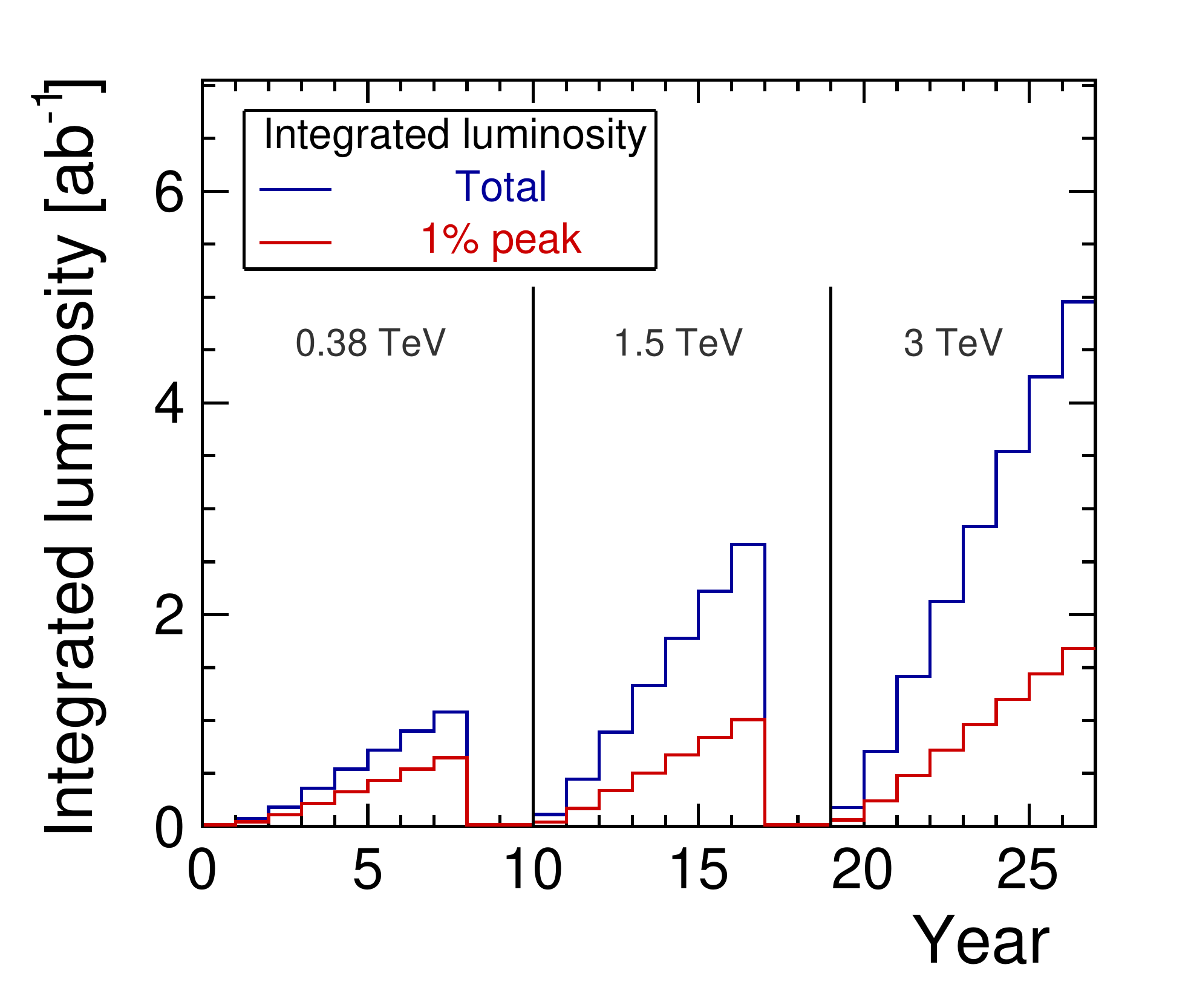}
\caption{Integrated luminosity in the updated CLIC staging scenario.  The luminosity ramp-up corresponds
         to what is described in \autoref{fig:clicstaging1}. \label{fig:clicstaging2}}
\end{center}
\end{figure}

The total integrated luminosities are summarized in \autoref{tab:clicstaging}. 
The ramp-up scenario is shown as luminosity per year in \autoref{fig:clicstaging1} and
total integrated luminosity in \autoref{fig:clicstaging2}.

\begin{table*}[htp]\centering
  \begin{tabular}{ccc|cc}\toprule
    & & & $P(\Pem)$=--80\% & $P(\Pem)$=+80\% \\
   Stage & $\sqrt{s}$ [TeV] & $\mathcal{L}_{\textrm int}$ [ab$^{-1}$] & $\mathcal{L}_{\textrm int}$ [ab$^{-1}$] & $\mathcal{L}_{\textrm int}$ [ab$^{-1}$] \\
   \hline
   1 &  0.38 (and 0.35) &  1.0 & 0.5 & 0.5 \\
   2 &  1.5             &  2.5 & 2.0 & 0.5 \\
   3 &  3.0             &  5.0 & 4.0 & 1.0 \\
    \bottomrule
  \end{tabular}
    \caption{Baseline CLIC energy stages and integrated luminosities for each stage in the updated scenario. \label{tab:clicstaging}}
\end{table*}

\section{Polarisation}
\label{sec:polarisation}

The CLIC baseline has $\pm 80$\% longitudinal polarisation for the electron beam, and no positron polarisation.
At the initial energy stage, the dominant Higgs production cross section is Higgsstrahlung,
which is not much affected by the electron polarisation.
Collecting data from both polarisations improves the sensitivity to certain 
BSM effects, for example in two-fermion production.  Consequently, 
equal amounts of --80\% and +80\% polarisation running are foreseen at the initial energy stage.
At the higher-energy stages, the dominant Higgs production mechanism is through $\PW\PW$ fusion,
and double-Higgs production becomes accessible. 
Single and double-Higgs production through $\PW\PW$-fusion is significantly enhanced
by around 80\% by running with --80\% electron polarisation, while 
it is reduced by the same amount with +80\% polarisation \cite{Abramowicz:2016zbo}. 
However, some +80\% electron polarisation running is required for sensitivity to certain BSM effects. 
For the two higher-energy stages, a baseline is adopted of sharing the running
time for --80\% and +80\% electron polarisation in the ratio 80:20. 

\section{Higgs couplings}
\label{sec:higgs}

One of the unique capabilities of lepton colliders is to measure the total production
cross section for the Higgsstrahlung process, $\sigma(\PZ\PH)$, using only the
system that recoils against the produced Higgs boson and without examining the Higgs
decay products.
In turn, this allows the Higgs couplings to be extracted in a model-independent
way (for example, without including any assumption about the absence of BSM invisible
decays of the Higgs boson) from the cross sections times branching ratios of individual 
final states.  The uncertainty of the measured $\sigma(\PZ\PH)$ therefore contributes
to the precision of all of the other Higgs coupling measurements.  
This recoil measurement is only possible at the initial CLIC energy stage,
and obtaining a high-accuracy measurement of $\sigma(\PZ\PH)$ is a main
motivation for planning a relatively long run at $\roots=380\,\GeV$. 

At the highest energy stage, several flagship measurements such as double Higgs boson
production are limited by statistics, providing a motivation for increased luminosity 
also at the higher energies.

\subsection{Summary of Higgs observables}

Extensive studies of the CLIC sensitivities to Higgs couplings have been reported
previously in \cite{Abramowicz:2016zbo}.
Details of the analyses and combined fitting can be found there.  
For those studies, an earlier energy staging
of $\sqrt{s}=350\,\GeV$, 1.4\,TeV, and 3\,TeV was assumed,
with corresponding integrated luminosities of 0.5, 1.5, and 3\,ab$^{-1}$.
As a consequence of those and other studies, the present initial stage energy of $\roots=380\,\GeV$
was adopted in order to optimise the physics reach of the initial stage by
including precise measurements of top-quark properties above threshold. 
In this document, the precisions of the individual Higgs sector measurements are updated for the
integrated luminosities of the present staging scenario: 1.0, 2.5, and 5.0\,ab$^{-1}$,
and for the polarisation baseline scenario, while still using the previous energy stages. 

Precisions on the Higgs observables are given in \autoref{tab:GlobalFit:Input350} for
the first energy stage, and in \autoref{tab:GlobalFit:Input143} for the two
higher-energy stages.  These individual results assume unpolarised beams.

Measurement of the cross section for double-Higgs production at 1.4 and 3\,TeV 
gives sensitivity to the Higgs self-coupling $\lambda$.
Assuming unpolarised beams, 
the precision on $\sigma(\PH\PH\PGne\PAGne)$ that comes from 2.5\,ab$^{-1}$ of integrated luminosity at 1.4\,TeV
results in a statistical precision on $\lambda$ of 42\%, which reduces to 18\% with 5.0\,ab$^{-1}$ at 3\,TeV.
Assuming $-80\%$ electron polarisation, these precisions improve to 31\% and 14\%, respectively.
The additional inclusion of differential measurements at 3\,TeV,
and the $\PZ\PH\PH$ double-Higgsstrahlung cross section at 1.4\,TeV, 
yields a precision of $[-7\%,+11\%]$ on $\lambda$.

The recoil mass analysis from $\Pep\Pem\to\PZ\PH$
events can be used to search for BSM decay modes of the Higgs boson into 
`invisible' final states.  Scaling the result from \cite{Abramowicz:2016zbo}
to the updated luminosity staging scenario gives an upper limit
on the invisible Higgs branching ratio of BR($\PH\to$ invis.) $<$ 0.69\%
at 90\% C.L. in the modified frequentist approach 
(for 1\,ab$^{-1}$ at $\roots=350\,\GeV$).

\begin{table*}[htp]\centering
  \begin{tabular}{lllcc}\toprule
                        &                                                           &                              & \tabt{Statistical precision} & \\ \cmidrule(l){4-4}
        \tabt{Channel}  & \tabt{Measurement}                                        & \tabt{Observable}            & $350\,\GeV$ & Reference \\ 
                        &                                                           &                              & $1\,\abinv$ & \cite{Abramowicz:2016zbo} \\ \midrule
    $\PZ\PH$            & Recoil mass distribution                                  & $\mH$                        & $78\,\MeV$ & \cite{Abramowicz:2016zbo} \\
    $\PZ\PH$            & $\sigma(\PZ\PH)\times \BR(\PH\to\text{invisible})$        & $\Gamma_\text{inv}$           & $0.4\,\%$ & \cite{Abramowicz:2016zbo} \\ \midrule
    $\PZ\PH$            & $\sigma(\PZ\PH)\times \BR(\PZ\to\Plp\Plm)$                & $\gHZZ^{2}$                   & $2.7\,\%$ & \cite{Abramowicz:2016zbo} \\
    $\PZ\PH$            & $\sigma(\PZ\PH)\times \BR(\PZ\to\PQq\PAQq)$               & $\gHZZ^{2}$                   & $1.3\,\%$ & \cite{Abramowicz:2016zbo} \\
    $\PZ\PH$            & $\sigma(\PZ\PH)\times \BR(\PH\to\PQb\PAQb)$               & $\gHZZ^{2}\gHbb^{2}/\GH$       & $0.61\,\%$ & \cite{Abramowicz:2016zbo} \\
    $\PZ\PH$            & $\sigma(\PZ\PH)\times \BR(\PH\to\PQc\PAQc)$               & $\gHZZ^{2}\gHcc^2/\GH$        & $10\,\%$ & \cite{Abramowicz:2016zbo} \\
    $\PZ\PH$            & $\sigma(\PZ\PH)\times \BR(\PH\to\Pg\Pg)$                  &                              & $4.3\,\%$ & \cite{Abramowicz:2016zbo} \\
    $\PZ\PH$            & $\sigma(\PZ\PH)\times \BR(\PH\to\tptm)$                   & $\gHZZ^{2}\gHTauTau^{2}/\GH$  & $4.4\,\%$ & \cite{Abramowicz:2016zbo} \\
    $\PZ\PH$            & $\sigma(\PZ\PH)\times \BR(\PH\to\PW\PW^*)$                & $\gHZZ^{2}\gHWW^{2}/\GH$      & $3.6\,\%$ & \cite{Abramowicz:2016zbo} \\
    $\PH\PGne\PAGne$    & $\sigma(\PH\PGne\PAGne)\times \BR(\PH\to\PQb\PAQb)$       & $\gHWW^{2}\gHbb^{2}/\GH$      & $1.3\,\%$ & \cite{Abramowicz:2016zbo} \\
    $\PH\PGne\PAGne$    & $\sigma(\PH\PGne\PAGne)\times \BR(\PH\to\PQc\PAQc)$       & $\gHWW^{2}\gHcc^{2}/\GH$      & $18\,\%$ & \cite{Abramowicz:2016zbo} \\
    $\PH\PGne\PAGne$    & $\sigma(\PH\PGne\PAGne)\times \BR(\PH\to\Pg\Pg)$          &                             & $7.2\,\%$ & \cite{Abramowicz:2016zbo} \\    
    \bottomrule
  \end{tabular}
    \caption{Summary of the precisions obtainable for the Higgs
      observables in the first stage of CLIC for an integrated
      luminosity of $1\,\abinv$ at $\roots=350\,\GeV$, assuming
      unpolarised beams. For the branching ratios, the measurement
      precision refers to the expected statistical uncertainty on the
      product of the relevant cross section and branching ratio; this
      is equivalent to the expected statistical uncertainty of the
      product of couplings divided by $\Gamma_{\PH}$ as indicated in
      the third column. \label{tab:GlobalFit:Input350}}
\end{table*}

\begin{table*}[htp]\centering
    \begin{tabular}{lllccc}\toprule
                        &                                                            &                             & \tabtt{Statistical precision} & \\ \cmidrule(l){4-5}
        \tabt{Channel}  & \tabt{Measurement}                                         & \tabt{Observable}           & $1.4\,\TeV$        & $3\,\TeV$ & Reference \\ 
                        &                                                            &                             & $2.5\,\abinv$      & $5.0\,\abinv$ & \\ \midrule
   $\PH\PGne\PAGne$     & $\PH\to\PQb\PAQb$ mass distribution                        & $\mH$                       & $36\,\MeV$         & $28\,\MeV$ & \cite{Abramowicz:2016zbo} \\ \midrule
   $\PZ\PH$             & $\sigma(\PZ\PH)\times \BR(\PH\to\PQb\PAQb)$                & $\gHZZ^{2}\gHbb^{2}/\GH$     & $2.6\,\%^{\dagger}$   & $4.3\,\%^{\dagger}$ & \cite{Ellis:2017kfi} \\
   $\PH\PGne\PAGne$     & $\sigma(\PH\PGne\PAGne)\times \BR(\PH\to\PQb\PAQb)$        & $\gHWW^{2}\gHbb^{2}/\GH$     & $0.3\,\%$          & $0.2\,\%$ & \cite{Abramowicz:2016zbo} \\
   $\PH\PGne\PAGne$     & $\sigma(\PH\PGne\PAGne)\times \BR(\PH\to\PQc\PAQc)$        & $\gHWW^{2}\gHcc^{2}/\GH$     & $4.7\,\%$          & $4.4\,\%$ & \cite{Abramowicz:2016zbo} \\
   $\PH\PGne\PAGne$     & $\sigma(\PH\PGne\PAGne)\times \BR(\PH\to\Pg\Pg)$           &                             & $3.9\,\%$          & $2.7\,\%$ & \cite{Abramowicz:2016zbo} \\
   $\PH\PGne\PAGne$     & $\sigma(\PH\PGne\PAGne)\times \BR(\PH\to\tptm)$            & $\gHWW^{2}\gHTauTau^{2}/\GH$  & $3.3\,\%$         & $2.8\,\%$ & \cite{Abramowicz:2016zbo} \\
   $\PH\PGne\PAGne$     & $\sigma(\PH\PGne\PAGne)\times \BR(\PH\to\mpmm)$            & $\gHWW^{2}\gHMuMu^{2}/\GH$    & $29\,\%$          & $16\,\%$ & \cite{Abramowicz:2016zbo} \\
   $\PH\PGne\PAGne$     & $\sigma(\PH\PGne\PAGne)\times \BR(\PH\to\upgamma\upgamma)$ &                             & $12\,\%$          & $6\,\%^*$ & \cite{Abramowicz:2016zbo} \\
   $\PH\PGne\PAGne$     & $\sigma(\PH\PGne\PAGne)\times \BR(\PH\to\PZ\upgamma)$      &                             & $33\,\%$          & $19\,\%^*$ & \cite{Abramowicz:2016zbo} \\
   $\PH\PGne\PAGne$     & $\sigma(\PH\PGne\PAGne)\times \BR(\PH\to\PW\PW^*)$         & $\gHWW^{4}/\GH$              & $0.8\,\%$         & $0.4\,\%^*$ & \cite{Abramowicz:2016zbo} \\
   $\PH\PGne\PAGne$     & $\sigma(\PH\PGne\PAGne)\times \BR(\PH\to\PZ\PZ^*)$         & $\gHWW^{2}\gHZZ^{2}/\GH$      & $4.3\,\%$         & $2.5\,\%^*$ & \cite{Abramowicz:2016zbo} \\
   $\PH\epem$           & $\sigma(\PH\epem)\times \BR(\PH\to\PQb\PAQb)$              & $\gHZZ^{2}\gHbb^{2}/\GH$      & $1.4\,\%$         & $1.5\,\%^*$ & \cite{Abramowicz:2016zbo} \\ \midrule
   $\PQt\PAQt\PH$       & $\sigma(\PQt\PAQt\PH)\times \BR(\PH\to\PQb\PAQb)$          & $\gHtt^{2}\gHbb^{2}/\GH$      & $5.7\,\%$         & $-$ & \cite{Abramowicz:2018rjq} \\
  \bottomrule
  \end{tabular}
  \caption{Summary of the precisions obtainable for the Higgs
    observables in the higher-energy CLIC stages for integrated
    luminosities of $2.5\,\abinv$ at $\roots=1.4\,\TeV$, and
    $5.0\,\abinv$ at $\roots=3\,\TeV$. In both cases unpolarised beams
    have been assumed. 
    For $\gHtt$, the $3\,\TeV$ case has not yet been studied. 
    Numbers marked with $*$ are extrapolated from $\roots=1.4\,\TeV$
    to $\roots=3\,\TeV$ while $\dagger$ indicates projections based on fast simulations.
    For the branching ratios, the measurement precision refers to the expected
    statistical uncertainty on the product of the relevant cross
    section and branching ratio; this is equivalent to the expected
    statistical uncertainty of the product of couplings divided by
    $\Gamma_{\PH}$, as indicated in the third column. \label{tab:GlobalFit:Input143}}
\end{table*}

\subsection{Combined fits}

Precisions on the Higgs couplings and width extracted from a model-independent
global fit, described in \cite{Abramowicz:2016zbo}, are given
in \autoref{tab:MIResultsPolarised8020} and \autoref{fig:MIResultsPolarised8020}.
The fit assumes the present baseline scenario for integrated luminosities and
beam polarisation.
The increase in cross section from having a predominantly negatively-polarised
electron beam is taken into account by multiplying the event rates for all
$\PW\PW$-fusion measurements by a factor of 1.48, corresponding to a factor of
1.8 for $80\,\%$ of the statistics and 0.2 for the remaining $20\,\%$. 
This approach is
conservative because it assumes that all backgrounds, including those from
$s$-channel processes, which do not receive the same polarisation enhancement,
scale by the same amount.

Each energy stage contributes significantly to the Higgs programme: 
the initial stage provides $\gHZZ$ and couplings to most fermions and
bosons, while the higher-energy stages improve them and add the top-quark,
muon, and photon couplings.  The precision on $\gHZZ$ is determined by the
statistics at the initial stage.

\begin{figure}
\begin{center}
\includegraphics[width=10cm]{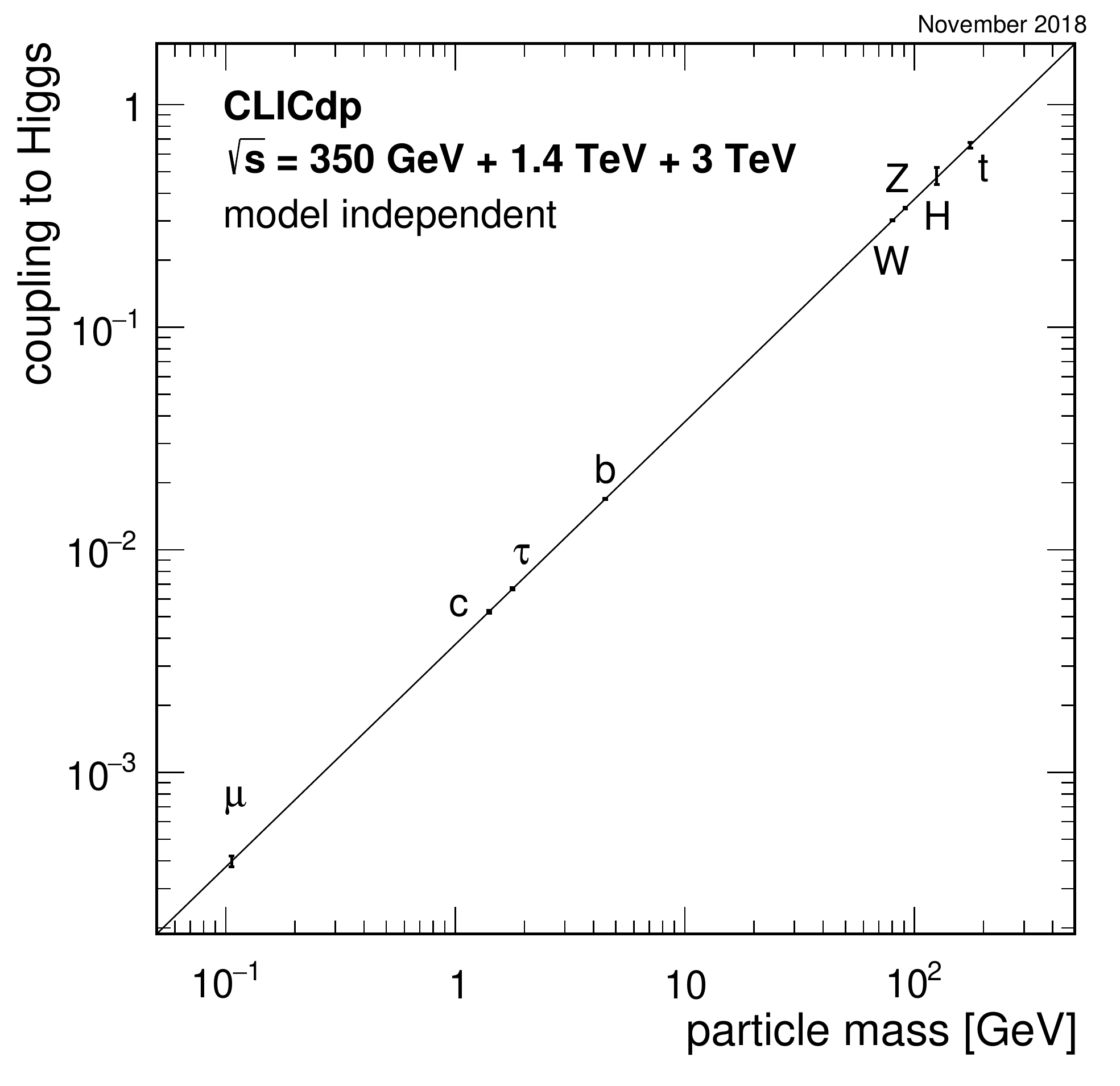}
\caption{Precision of the model-independent Higgs couplings and the self-coupling as a function of particle mass.
  The line shows the SM prediction that the Higgs coupling of each particle is proportional to its mass.\label{fig:couplingsMass}}
\end{center}
\end{figure}

\begin{figure}
\begin{center}
\includegraphics[width=12cm]{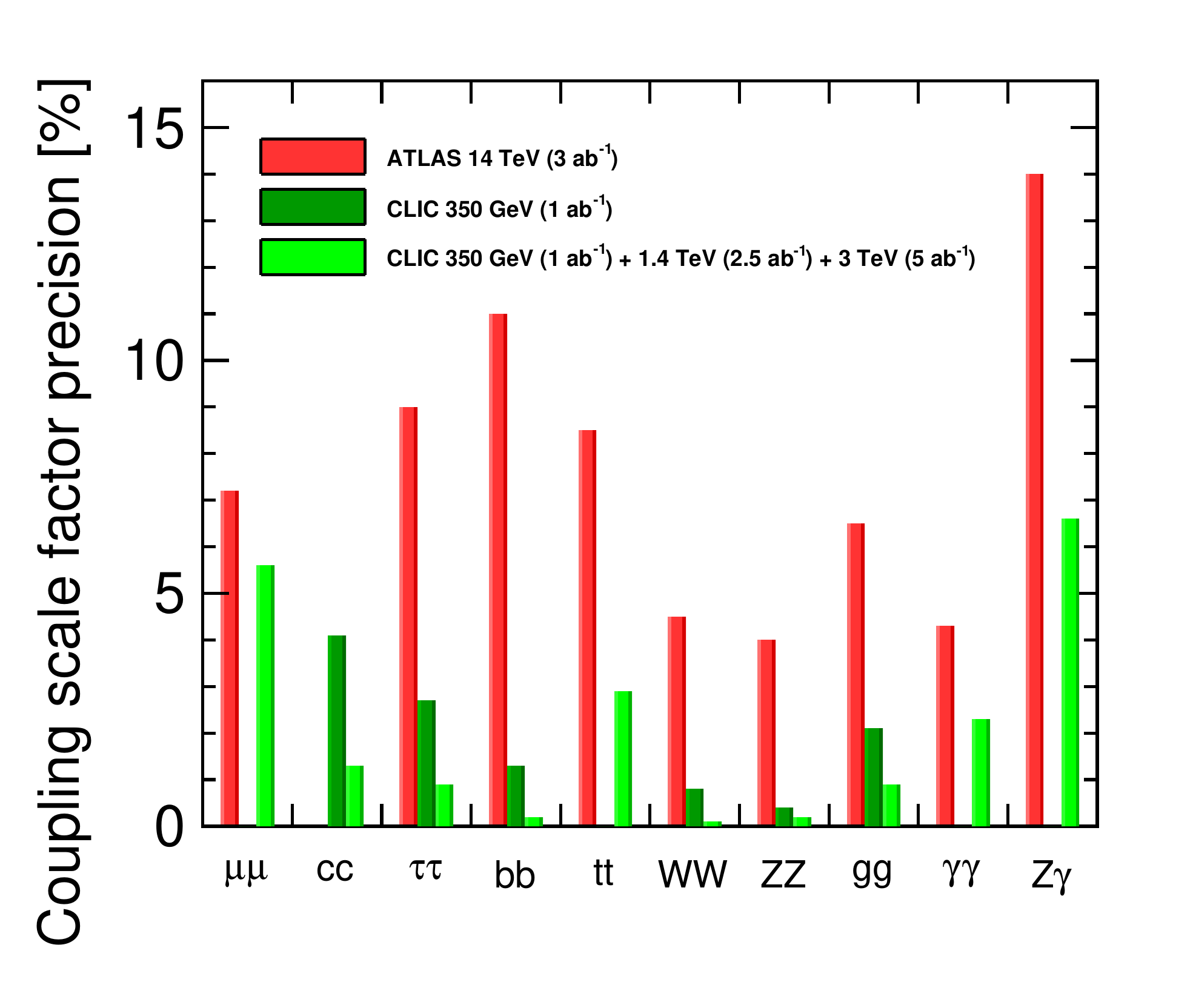}
\caption{Comparison of the model-dependent Higgs couplings with projections \cite{ATLAS_HLLHChiggs2014} for HL-LHC. \label{fig:hllhcComparison}}
\end{center}
\end{figure}

The final precision of the model-independent couplings are shown as a function of particle mass in \autoref{fig:couplingsMass}.
Precisions extracted from a model-dependent
global fit, also described in \cite{Abramowicz:2016zbo}, are given in \autoref{tab:MDResultsPolarised8020}
and \autoref{fig:MDResultsPolarised8020}.
This fit also assumes the present baseline scenario for integrated luminosities and
beam polarisation.
Comparisons between the CLIC-only model-dependent Higgs coupling sensitivities and corresponding
HL-LHC projections \cite{ATLAS_HLLHChiggs2014} are given in \autoref{fig:hllhcComparison}, showing 
that already after the initial energy stage, in many cases the CLIC precision is
significantly better than for the HL-LHC, and improves further with the higher-energy running.  

\section{Conclusions}

The updated CLIC luminosity staging baseline assumes 1.0, 2.5, and 5.0\,ab$^{-1}$ 
collected at the three energy stages of $\roots= 380\,\GeV, 1.5\,\TeV$, and $3\,\TeV$, respectively.

A baseline polarisation scenario foresees running time for
--80\% and +80\% electron polarisation shared equally at the initial energy stage,
and shared in the ratio 80:20 for the two higher-energy stages.

Updated Higgs coupling sensitivities are presented for this new baseline scenario 
that show excellent precision at the percent level for many of the Higgs couplings
from a model-independent fit, and precisions from a model-dependent fit that are
in many cases significantly better than those projected for the HL-LHC.

\clearpage

\begin{minipage}{\linewidth}
  \centering
  \begin{minipage}{0.495\textwidth}
    \begin{table}[H]
\begin{tabular}{lccc}
\toprule
Parameter & \multicolumn{3}{c}{Relative precision}\\
\midrule
& $350\,\GeV$ & + $1.4\,\TeV$ & + $3\,\TeV$\\
&$1\,\abinv$& + $2.5\,\abinv$& + $5\,\abinv$\\
\midrule
$\gHZZ$ & 0.6\,\% & 0.6\,\% & 0.6\,\% \\
$\gHWW$ & 1.0\,\% & 0.6\,\% & 0.6\,\% \\
$\gHbb$ & 2.1\,\% & 0.7\,\% & 0.7\,\% \\
$\gHcc$ & 4.4\,\% & 1.9\,\% & 1.4\,\% \\
$\gHTauTau$ & 3.1\,\% & 1.4\,\% & 1.0\,\% \\
$\gHMuMu$ & $-$ & 12.1\,\% & 5.7\,\% \\
$\gHtt$ & $-$ & 3.0\,\% & 3.0\,\% \\
\midrule
$g^\dagger_{\PH\Pg\Pg}$ & 2.6\,\% & 1.4\,\% & 1.0\,\% \\
$g^\dagger_{\PH\PGg\PGg}$ & $-$ & 4.8\,\% & 2.3\,\% \\
$g^\dagger_{\PH\PZ\PGg}$ & $-$ & 13.3\,\% & 6.7\,\% \\
\midrule
$\Gamma_{\PH}$ & 4.7\,\% & 2.6\,\% & 2.5\,\% \\
\bottomrule
      \end{tabular}
      \caption{ }\label{tab:MIResultsPolarised8020}
    \end{table}
  \end{minipage}
  \begin{minipage}{0.495\textwidth}
    \begin{figure}[H]
      \includegraphics[width=\linewidth]{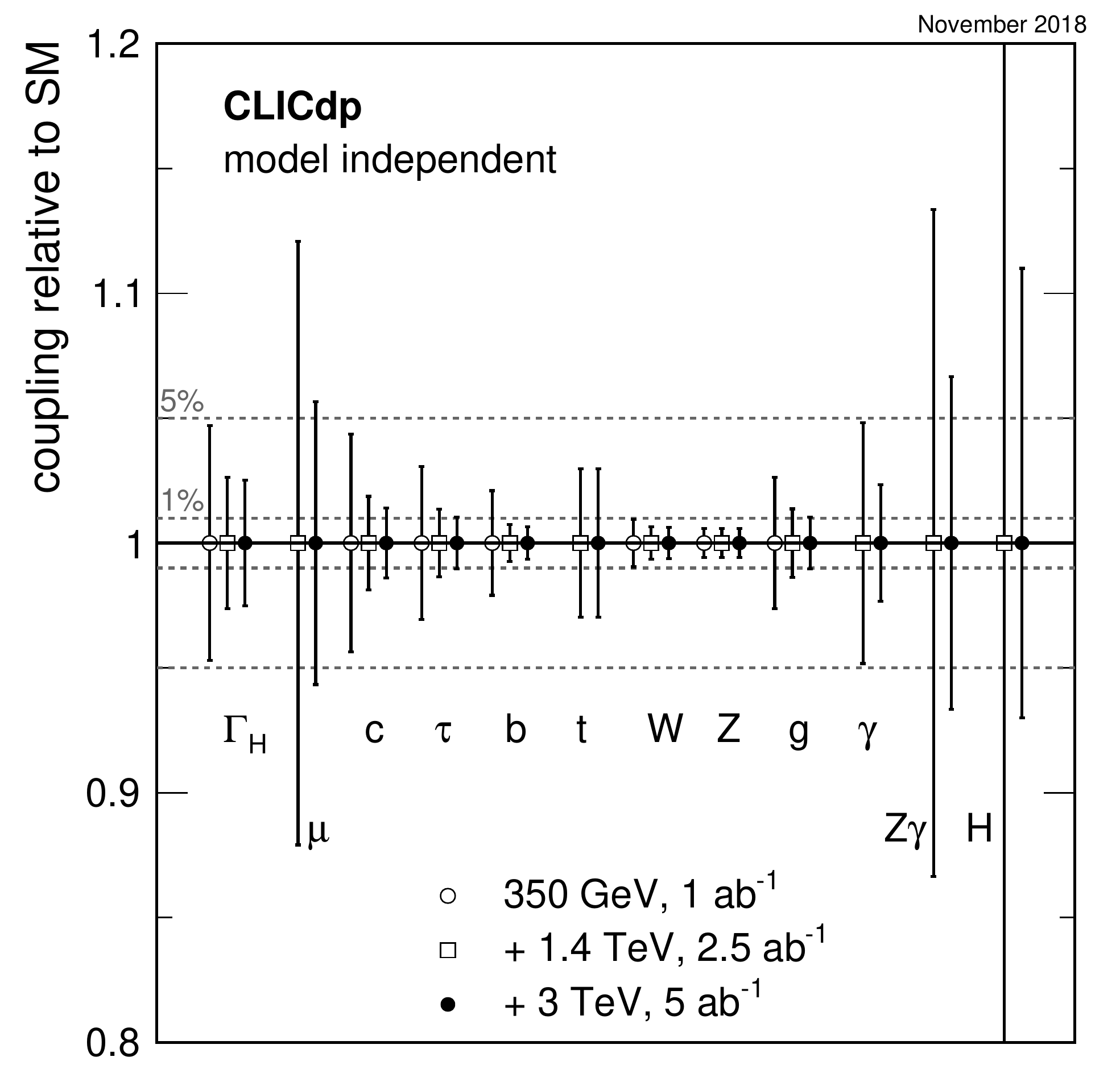}
      \caption{ }\label{fig:MIResultsPolarised8020}
    \end{figure}
  \end{minipage}
        Results of the model-independent fit.
        For $\gHtt$, the $3\,\TeV$ case has not yet been studied. The three
        effective couplings $g^\dagger_{\PH\Pg\Pg}$, 
        $g^\dagger_{\PH\PGg\PGg}$ and $g^\dagger_{\PH\PZ\PGg}$ are also included in the fit. 
        Operation with $-80\,\%$ ($+80\,\%$) electron beam polarisation is assumed for 
        $80\,\%$ ($20\,\%$) of the collected luminosity above 1\,\TeV, corresponding
        to the baseline scenario. 
\end{minipage}

\begin{minipage}{\linewidth}
\vspace*{1cm}
  \centering
  \begin{minipage}{0.495\textwidth}
    \begin{table}[H]
\begin{tabular}{lccc}
\toprule
Parameter & \multicolumn{3}{c}{Relative precision}\\
\midrule
& $350\,\GeV$ & + $1.4\,\TeV$& + $3\,\TeV$\\
&$1\,\abinv$& + $2.5\,\abinv$& + $5\,\abinv$\\
\midrule
$\kappa_{\PH\PZ\PZ}$ & 0.4\,\% & 0.3\,\% & 0.2\,\% \\
$\kappa_{\PH\PW\PW}$ & 0.8\,\% & 0.2\,\% & 0.1\,\% \\
$\kappa_{\PH\PQb\PQb}$ & 1.3\,\% & 0.3\,\% & 0.2\,\% \\
$\kappa_{\PH\PQc\PQc}$ & 4.1\,\% & 1.8\,\% & 1.3\,\% \\
$\kappa_{\PH\PGt\PGt}$ & 2.7\,\% & 1.2\,\% & 0.9\,\% \\
$\kappa_{\PH\PGm\PGm}$ & $-$ & 12.1\,\% & 5.6\,\% \\
$\kappa_{\PH\PQt\PQt}$ & $-$ & 2.9\,\% & 2.9\,\% \\
$\kappa_{\PH\Pg\Pg}$ & 2.1\,\% & 1.2\,\% & 0.9\,\% \\
$\kappa_{\PH\PGg\PGg}$ & $-$ & 4.8\,\% & 2.3\,\% \\
$\kappa_{\PH\PZ\PGg}$ & $-$ & 13.3\,\% & 6.6\,\% \\
\bottomrule
\end{tabular}
\caption{ }\label{tab:MDResultsPolarised8020}
    \end{table}
  \end{minipage}
  \begin{minipage}{0.495\textwidth}
    \begin{figure}[H]
      \includegraphics[width=\linewidth]{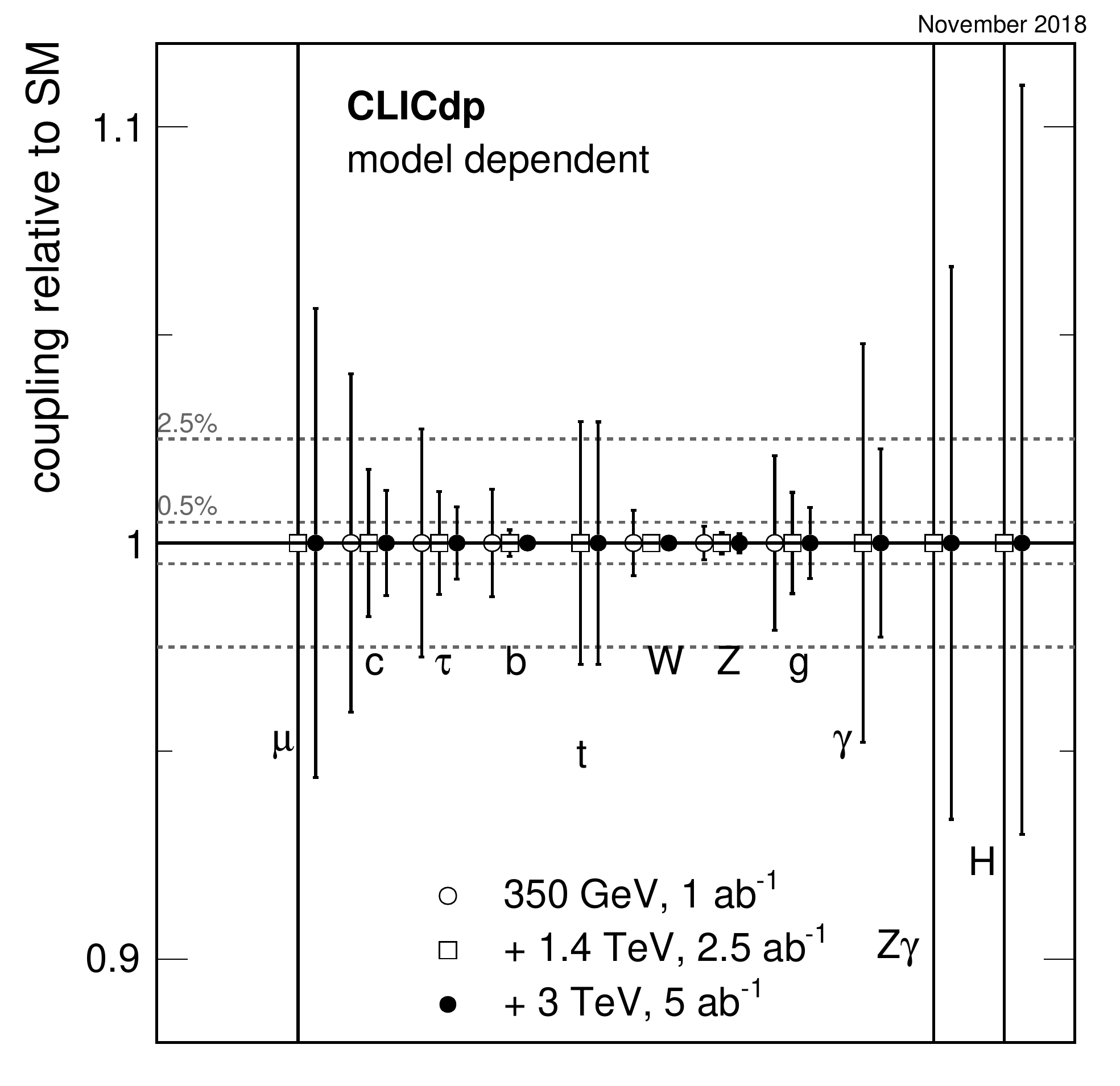}
      \caption{ }\label{fig:MDResultsPolarised8020}
    \end{figure}
  \end{minipage}
  Results of the model-dependent fit without theoretical uncertainties. 
  For $\kappa_{\PH\PQt\PQt}$, the $3\,\TeV$ case has not yet been
  studied. The uncertainty of the total width is calculated from the fit results. 
  Operation with $-80\,\%$ ($+80\,\%$) electron beam polarisation is assumed for 
  $80\,\%$ ($20\,\%$) of the collected luminosity above 1\,\TeV, corresponding to the baseline scenario. 
\end{minipage}

\printbibliography[title=References]

\end{document}